\documentstyle[twocolumn,aps,pra,epsfig,floats]{revtex}
\newcommand{\QKD}{{\sc qkd}}

\newcommand{\CPS}{{\sc cps}}
\newcommand{\PNR}{{\sc pnr}}
\newcommand{\WCP}{{\sc wcp}}

\newcommand{\SPDC}{{\sc spdc}}

\newcommand{\CPR}{\CPS/\PNR}
\newcommand{\G}{G}
\newcommand{\pexp}{p_{exp}}
\newcommand{\ps}{p_{s}}
\newcommand{\ebar}{\overline{\epsilon}}
\newcommand{\Sm}{S_m}
\newcommand{\rone}{R_1}

\begin{document}
\twocolumn
\wideabs{
\title{Performance of Photon-Pair Quantum Key Distribution Systems}
\author{Z.~Walton,$^1$ A.\,V.~Sergienko,$^{1,2}$ M.~Atat\"{u}re,$^2$ B.\,E.\,A.~Saleh,$^1$ and M.\,C.~Teich$^{1,2}$}
\address{$^1$Quantum Imaging Laboratory, Department of Electrical and Computer Engineering, Boston University,
8 Saint Mary's Street, Boston, Massachusetts 02215\\
$^2$Department of Physics, Boston University, 8 Saint Mary's Street, Boston, Massachusetts 02215}

\date{\today}
\maketitle
\begin{abstract}We analyze the quantitative improvement in performance provided by a novel quantum key distribution (\QKD) system that employs a correlated
photon source (\CPS) and a photon-number resolving detector (\PNR).
Our calculations suggest that given current technology, the \CPR\
implementation offers an improvement of several orders of magnitude in
secure bit rate over previously described implementations.

%
\end{abstract}
}

\section{Introduction}

While much progress has been made in the field of experimental quantum
key distribution (\QKD) since the first proof-of-principle in
1992~\cite{bennett}, the failure of the experimental community to
choose a well-defined scope for the technological power of the
eavesdropper has made comparing the competing implementations
difficult.  Specifically, the mean number of photons per pulse is
arbitrarily set at approximately $0.1$ photons per pulse by
most groups. There are two problems with operating the source
at this power.  First, since this mean value is not determined by
maximizing the appropriate figure of merit (i.e., secure bits per pulse), each implementation must be assumed to be operated at a
sub-optimal point in the parameter space, making it difficult to quantify the
performance advantage enjoyed by one system over another.  Second,
recent work has shown that the choice of $0.1$ photons per pulse makes
all existing weak coherent pulse implementations insecure to an
eavesdropper armed with foreseeable, though not presently available,
technology~\cite{brassard}.

In this paper, we combine reported experimental results in the
literature with a specific scope for the eavesdropper and
L\"utkenhaus' fully secure version~\cite{lutkenhaus} of the BB84
protocol~\cite{bb84} to determine which of three physical
implementations provides the best performance for free-space and
optical fiber applications.  The first two implementations, based on
weak coherent pulses (\WCP) and correlated photon sources (\CPS)
respectively, have been investigated elsewhere~\cite{brassard}; the
third implementation (\CPR) is a new design that combines the perfect
photon-number correlation in spontaneous down
conversion~\cite{sergienko} with photon-number resolving detectors
(\PNR)~\cite{teich-pnr,vlpc} to reduce the effect of the multi-photon
security loophole. Our calculations indicate that this novel design
offers a substantial advantage over the competing implementations,
mainly because of its closer approximation to the true single-photon
state.

Most reports of the performance of specific \QKD\ systems either
ignore the vulnerability of the system to eavesdropper attack or
provide special-case analyses in which the information accessible to
an eavesdropper employing a specific attack is estimated.  This runs
counter to the fundamental paradigm of quantum cryptography.  While
conventional public-key cryptosystems are based on unproven propositions of
theoretical computer science and can only be used against an adversary
who has limited computational power, quantum cryptography promises
unconditional security {\em regardless} of the technological
capabilities of the adversary.  Thus, candidate \QKD\ systems should
be evaluated in this context.

Our analysis places no technological limitations on the eavesdropper
(Eve) except that she attacks each pulse individually.
Although it is not yet proven, it it widely believed that restricting
Eve to individual attacks does not prevent her from performing the
optimal attack.  The essence of the argument is that Eve's techniques
for learning information about any two pulses are in no way
restricted by requiring her to gain information from each separately,
since the two parties (Alice and Bob) are attempting to share a random bit string in
which any two bits are completely uncorrelated.

\section{The Figure of Merit: Secure Bits per Pulse}

The existence of classical privacy amplification algorithms for
distilling arbitrarily secure bits from partially secure bits means
that it is not necessary to have complete security for each pulse. As
long as a bound on the information leaked to the adversary
can be inferred from measurable quantities, such as the observed error
rate, Alice and Bob can recover a perfectly secure, shared key by a
two-step procedure.  They first use traditional error-correcting
methods to ensure they have the same key, and then use the technique
of generalized privacy amplification~\cite{bennett2} to extract a
shorter secure key from a longer key.  Thus, the
crucial figure of merit for \QKD\ implementations is the fraction of
the raw bits shared by Alice and Bob that may be kept, such that they
are certain that they share the same key and that Eve has
negligible information about that key.

This fraction, labeled $\G$ for gain, depends on four factors: the observed
error rate ($\ebar$), the probability that Alice's detector-triggered
source indicates that a valid signal was created ($\ps$), the
probability that Alice sends a multi-photon pulse ($\Sm$), and the
probability that a pulse sent by Alice leads to a successful detection
by Bob ($\pexp$).  The dependence of $\G$ on $\ebar$ for the BB84
protocol faced with the aforementioned adversary was determined by
C.~Fuchs et al. in 1997~\cite{fuchs}; however, the more crucial
dependence of $\G$ on $\ps$, $\Sm$, and $\pexp$ has only recently been
determined by L\"utkenhaus~\cite{lutkenhaus}.  Combining these
two analyses, we have
\begin{eqnarray*}
& & \G(\ebar,\ps,\Sm,\pexp) \\
&=& \frac{1}{2}\ps\pexp\left\{-\rone\log_2\left[\frac{1}{2}+2\ebar\rone-2\left(\ebar\rone\right)^2\right]\right. \\
& & + 1.35[\ebar\log_2\ebar+(1-\ebar\log_2(1-\ebar)]\},
\end{eqnarray*}
where $\rone=\frac{\pexp-\Sm}{\pexp}$. It should be noted that for
this derivation of $\G$, the most conservative approach to
the imperfections in Bob's apparatus has been used: Eve has complete
control over all of the errors, photon losses, background, and dark
counts that occur in the optical channel {\em and} in Bob's detection
unit.  If it is assumed that Eve cannot control the imperfections 
in Bob's apparatus, the fraction $\G$ increases; however, it is
difficult to prove exactly which aspects of Bob's apparatus Eve may or
may not be able to influence. Thus, it seems prudent to assume the
worst case, as we have done here.

\section{Three QKD Source Designs}\label{designs}

A complete \QKD\ implementation consists of the physical apparatus and
a protocol which specifies how the apparatus should be operated, and
which provides probabilistic statements that characterize the outcome
(i.e., with probability $\epsilon$, Eve's guess at the secret key will be correct in more than half of the bits).
Since BB84 is the only protocol for which there exists an agreed-upon
method for calculating $\G(\ebar,\ps,\Sm,\pexp)$ in the face of
our adversary~\cite{lutkenhaus}, we use this protocol exclusively in
comparing the performance of the three implementations: \WCP, \CPS,
and \CPR.


The physical apparatus required for the BB84 protocol can be
conveniently partitioned into the single-photon source, the optical
channel, and the detection unit.  Several single-photon source
technologies are being considered for use in a complete \QKD\
system. Before presenting the results of our calculations, we
summarize the qualitative advantages and disadvantages of the three
leading single-photon-source technologies.

\begin{figure}[t!]
    \vspace {-0.3 cm}
\begin{center}
\epsfxsize=\columnwidth
\epsfbox{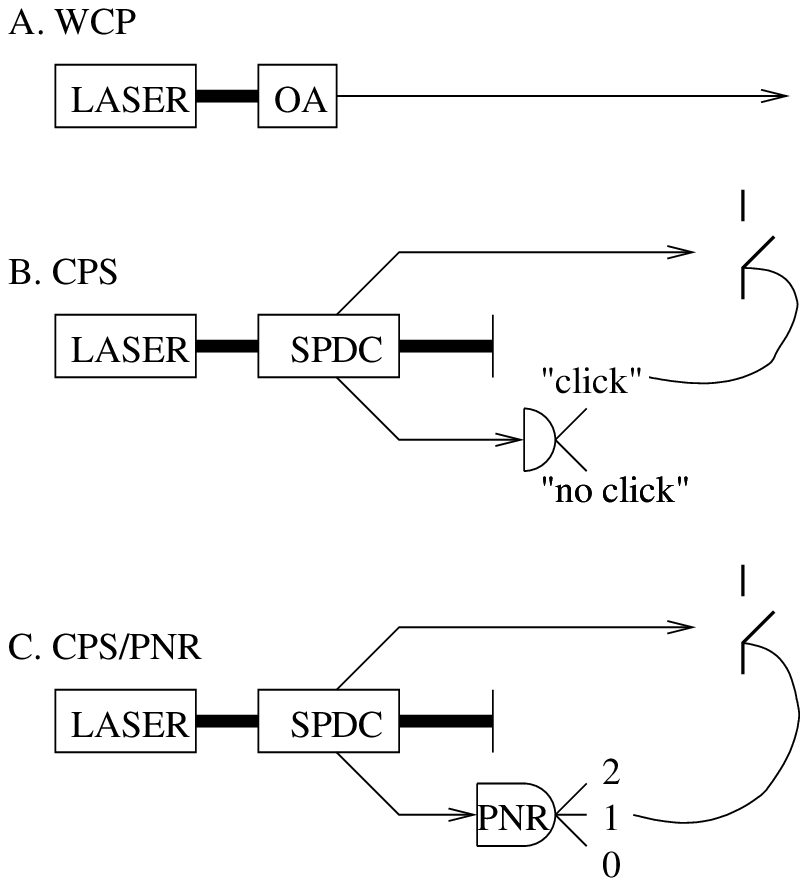}
\end{center}
\caption{{\bf Three QKD source designs.}  In
A, a weak coherent pulse (\WCP) from a laser source is optically attenuated (OA) to a mean photon number much less then one (the polarization rotator necessary for
implementing the BB84 protocol is not shown).  Both B and C are
detector-triggered sources based on spontaneous parametric down
conversion (\SPDC) in which Alice only allows the pulse in the signal beam to
propagate to Bob if her detector indicates that one photon arrived in
the idler beam.  In B the idler beam is monitored with a standard
``click''/``no click'' detector.  In C, the idler beam is monitored
with a photon-number resolving detector (\PNR), which
can discriminate between single- and double-photon arrivals.  By not
using the pulses that she determines contain multiple photons,
Alice significantly improves the secure bit rate and extends the range
of tolerable channel loss.}
\label{imps}
\end{figure}

\subsection{Weak Coherent Pulse (WCP)}

The simplest and most common method of reducing the probability of
a multi-photon pulse is to attenuate a weak coherent pulse (\WCP) of
light from a laser (see Fig.~\ref{imps}A).  Since a partitioned
Poisson random variable still exhibits Poisson statistics, Alice must
adjust the mean photon number per pulse in order to strike a balance
between two undesirable effects: the wasteful zero-photon pulses and
the insecure multi-photon pulses. Once the pulse is created, Alice
and Bob may use standard optical components to modify, launch,
transmit, collect and measure the polarization of the optical pulse.
Since the different sources we consider work equally well with
the other parts of the complete \QKD\ apparatus, we leave these
aspects of the apparatus unspecified and base our calculations on
values for optical coupling efficiency, error probabilities, and
detector performance reported in the
literature~\cite{marand,townsend,buttler}.

\subsection{Correlated Photon Source (CPS)}

In the paper that reveals the complete insecurity of current \WCP\
implementations~\cite{brassard}, Brassard et al.~investigate the
ability of a detector-triggered source based on spontaneous parametric
down conversion (\SPDC) to mitigate the multi-photon security loophole
(see Fig.~\ref{imps}B).  The perfect correlation in photon number in
the signal and idler beams allows Alice to run the protocol only when
her detectors on the idler beam indicate that one photon was sent to
Bob along the signal beam.  While this implementation of the
correlated photon source (\CPS) extends the range of permissible
channel losses several orders of magnitude from that allowed in the
\WCP\ case~\cite{lutkenhaus,brassard}, the Poisson statistics for the
number of pairs per pulse~\cite{teich-poisson} combined with the
inability of standard detectors to distinguish single- and
multi-photon detection events lead to a persistence of the insecure
multi-photon pulses.

\subsection{Correlated Photon Source with a Photon-Number Resolving Detector (CPS/PNR)}

To minimize the chance that Alice registers a valid signal when more
than one pair was created, we place a photon-number resolving detector
in Alice's laboratory.  In our calculations we use the characteristics
of the photon-number resolving detector reported in Ref.~\cite{vlpc},
since this device is representative of the state-of-the-art. While
this detector has a finite quantum efficiency of approximately 70\%,
the gain mechanism ensures that the device can distinguish the number
of photoelectron-multiplication events with very low error
($\sim$0.63\%).  The relatively high dark count rate ($\sim$10$^4$
counts per second) can be effectively mitigated by limiting the
detector's exposure time by nanosecond gating. By initiating a pulse
transmission only when the detector reports one photon arriving, Alice
significantly reduces the fraction of pulses sent to Bob that contain
more than one photon.

The difficulties with this approach stem from the extreme conditions
necessary for the \PNR\ to provide such high efficiency and low
multiplication noise.  The actively controlled, bath-type He cryostat
required for optimal performance~\cite{vlpc} precludes miniaturization of the
source and complicates the task of creating a \QKD\
implementation that is reliable, durable and economically feasibly for
real-world applications.  Nonetheless, our simulations indicate that,
in achieving a closer approximation to the true single-photon source,
the \CPR\ implementation provides an option for obtaining a secure
link for certain applications in which existing implementations provide
negligible gain.

\section{Examples}

We calculated the performance of the three implementations over both
free space and fiber-optic channels using values for optical coupling
efficiency, error probabilities, and detector performance reported in
the literature~\cite{marand,townsend,buttler}.  In each case the
performance was determined by maximizing $\G$ over the power of the
original laser pulses that are either attenuated (\WCP) or
down-converted (\CPS\ and \CPR) to create the pulse.  It is this
crucial step that most experimental groups have ignored, leading to mean
photon numbers that are orders of magnitude away from optimality and
to unrealistic claims concerning secure bit rates.  While the
experiments have been performed at specific distances, we extrapolate
the predicted gain over a range of distances by reasoning that the
dependence of $\G$ on distance is dominated by absorption in optical
fibers and diffraction in a free-space link. 

As graphed in Figs.~\ref{freespace} and~\ref{fibers}, each of the curves stops at a specific
distance along the x-axis and fails to descend off the bottom of the
plot, suggesting that there may be valid operating points with gain beyond the end of the curve.  It should be understood
that the true shape of these curves is nearly vertical at the
cut-off distance---the plot fails to convey this steep drop-off because
the numerical sampling algorithm used by the plotting program is not
fine enough to show the curves' continuity.

\subsection{Free-Space QKD}

\subsubsection{Ground-to-Ground Link}
\begin{figure}[t!]
\begin{center}
\epsfxsize=\columnwidth
\epsfbox{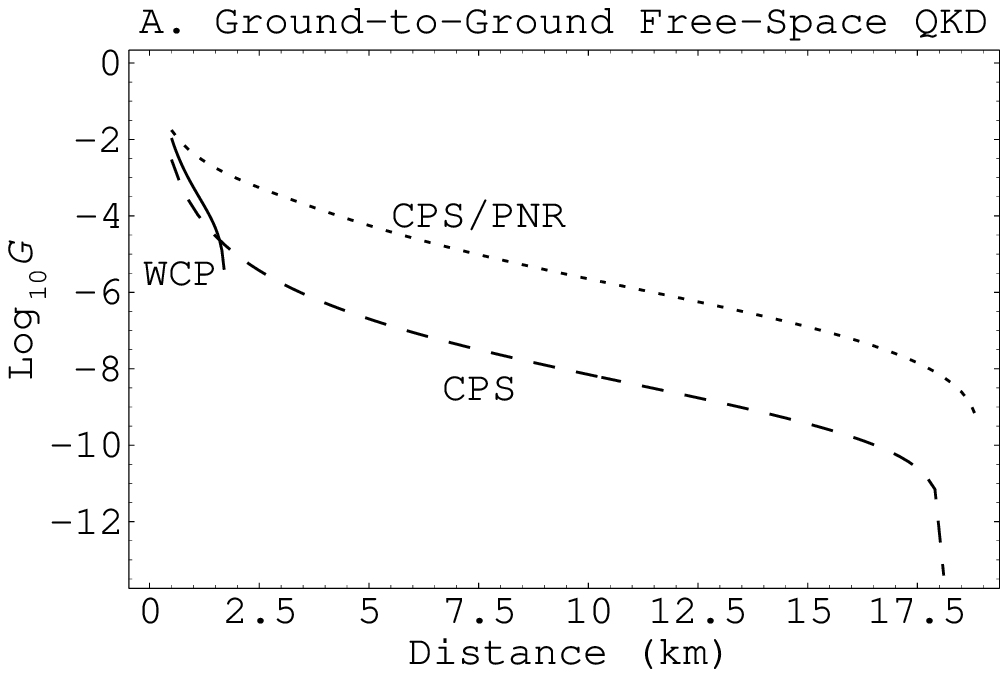}
\epsfxsize=\columnwidth
\epsfbox{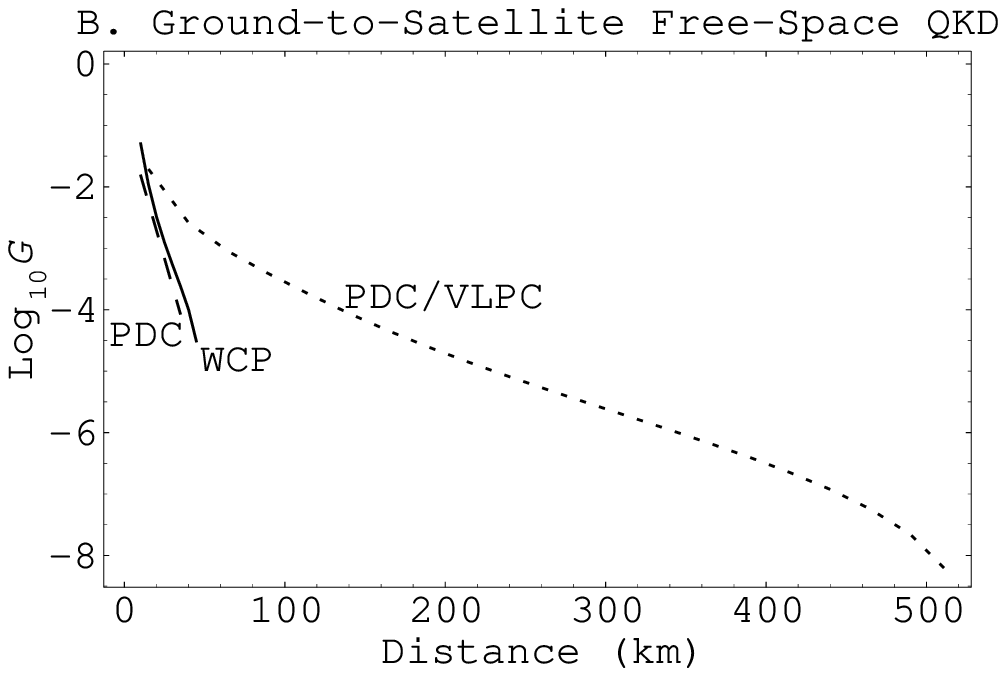}
\end{center}
\vspace {0 cm}
\caption{{\bf Free-space \QKD\ in ground-to-ground (A) and
ground-to-satellite (B) configurations for the three source
designs of Section~\protect\ref{designs}.} The gain \G\
represents the number of perfectly shared, secure bits, per pulse.
Note the difference scales in A and B. The values at 1 km in A and 300
km in B are based on the parameters for channel loss, error, and
background reported in Ref.~\protect\cite{buttler}.  The gain at all other
distances is calculated by assuming that the optical coupling
efficiency varies as $\frac{1}{d^2}$ as a result of beam diffraction, where
$d$ is the distance of the transmission.}
\label{freespace}
\end{figure}

Fig.~\ref{freespace}A shows the relative performance of the three
implementations along the surface of the Earth under nighttime
conditions.  The values of the gain at $d=1$ km (i.e.,
\WCP$\rightarrow5.6\times 10^{-4}$, \CPS$\rightarrow1.5\times
10^{-4}$, and \CPR$\rightarrow4.2\times 10^{-3}$) represent the actual
values that could be achieved using the experimental apparatus
reported by Buttler et al.~for signal launch, collection and
detection~\cite{buttler}.  Thus, using a base repetition rate of 100
MHz, the \CPR\ implementation offers a 400 kbits/sec perfectly secure
channel.  The rate of this channel is approximately one order of
magnitude greater than that offered by the \WCP\ and \CPS\
implementations.  The most dramatic feature in Fig.~\ref{freespace}A
is the precipitous decline of the \WCP\ gain around 2 km.  The
persistence of the \CPS\ and \CPR\ curves beyond 10 km suggest that a
detector-triggered source would be required for secure communications
in a metropolitan area or battlefield, while the \WCP\ would be
sufficient for close proximity, building-to-building applications.

\subsubsection{Ground-to-Satellite Link}

In~\cite{buttler}, Buttler et al.~provide rough estimates of the
optical coupling efficiency and background rates in a
ground-to-satellite \QKD\ application.  Using these estimates, we have
simulated the gain achievable with each implementation for a range of
low-Earth orbit altitudes (see Fig.~\ref{freespace}B).  The apparent
discrepancy between Fig.~\ref{freespace}A and
Fig.~\ref{freespace}B---both describe free-space implementations, yet
Fig.~\ref{freespace}B shows gain far past the 20-km cutoff of
Fig.~\ref{freespace}A---is understood by observing that all but
the lowest 2 km of the ground-to-satellite link is turbulence-free
vacuum.  Our results indicate that while the \WCP\ and \CPS\
implementations offer no secure communication at standard low Earth
orbit altitudes ($\sim$100 km), the \CPR\ implementation could enable
the exchange of approximately $10^3$ secret bits for each nighttime
exchange (assuming a 10-MHz repetition rate, a 300-km orbit, and a
several-minute line-of-sight exposure between the ground station and
the satellite).


A complicating factor in these estimates is that the satellite
altitude determines the velocity necessary to remain in orbit.  While
a very low orbit would allow increased gain, the amount of time that
the satellite spends in sight of the ground station would be reduced,
decreasing the total number of secret bits shared in one pass.  It
seems reasonable to delay a determination of the optimal satellite
altitude until the exact characteristics of each element in the
proposed communication system are established.

\subsection{Optical Fiber QKD}

\begin{figure}[t!]  \begin{center} \epsfxsize=\columnwidth
\epsfbox{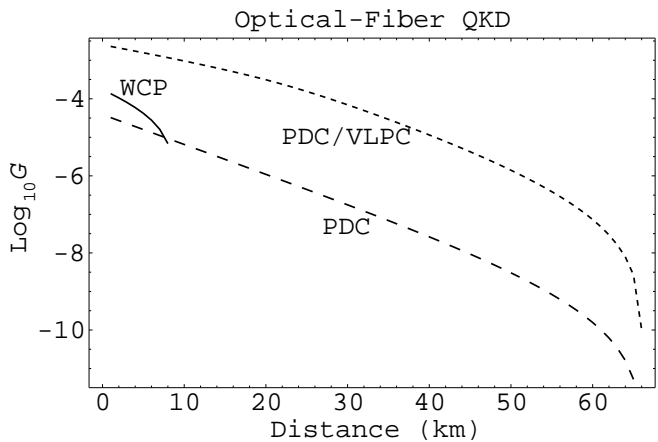} \end{center} \vspace {0 cm} \caption{{\bf Gain
through an optical fiber for the three source designs of Section~\protect\ref{designs}.} Transmission wavelength is set at
the first telecom window (1.3 $\mu$m) to achieve low loss (0.38 dB/km)
and to optimize detector performance (the detectors used had quantum
efficiency 0.11 and $10^{-5}$ dark counts per pulse duration).
Detector-triggered sources (i.e., \CPS\ and \CPR) use idler beams at
0.8 $\mu$m where detectors have higher efficiency and lower
noise. Calculations are based on the experimental values provided
in Refs.~\protect\cite{marand,townsend}. Note that the scales differ from those in Fig.~\protect\ref{freespace}}  \label{fibers} \end{figure}

Figure~\ref{fibers} confirms the conclusion of
Refs.~\cite{brassard,lutkenhaus}: the detector-triggered source
offers gain far beyond the $\sim$10 km cutoff distance of the
\WCP\ implementation through optical fiber.  Unlike these papers which
focused entirely on ``click''/``no click'' detectors in Alice's
source, our results indicate the considerable increase in gain offered
by photon-number resolving detectors.  Comparing Fig.~\ref{fibers} to
Fig.~\ref{freespace}A, it is clear that fiber-based \QKD\ offers
 performance superior to that of free-space \QKD, and is the obvious choice for
long-distance ground-to-ground applications, as a result of its immunity from diffraction,
background light, cloud cover, and temperature-dependent turbulence.
However, given current technology, ground-to-satellite free-space
\QKD\ with a \CPR\ source appears to be the preferred option for implementing a global, secure network.

\section{Discussion}

We have calculated the performance currently attainable with \QKD\
systems through free-space and optical fibers, for three different
 source designs, in the face of an unrestricted adversary who
attacks each pulse individually.  Our results indicate that the
implementation based on a correlated photon source (\CPS) offers the
best performance, as a result of the potentially unlimited precision in
identifying the presence of a single photon.  Furthermore, while using
a detector-triggered source extends the range of a \QKD\ system,
exploiting the photon-number resolving capabilities of a photon-number
resolving detector (\PNR) to decrease the fraction of multi-photon
pulses provides a further increase of several orders of magnitude
in $\G(\ebar,\ps,\Sm,\pexp)$, as seen in Figs.~\ref{freespace} and
\ref{fibers}.  We conclude that future progress in practical \QKD\ will
come largely from advances in detector performance and in the attendant
improvement in the detector-triggered single-photon source.

A summary of our calculations is as follows.  Using a base repetition
rate of 100 MHz for the pump laser, the \CPR\ implementation provides
a 400-kbits/sec secure channel over 1 km of free space, 100 bits/sec over 50 km of
optical fiber, and 100 bits/sec to a satellite in low earth orbit. The
two competing implementations provide at best only 50 kbits/sec over 1 km of free space,
1 bits/sec over 50 km of optical fiber, and cannot safely communicate with a
satellite at any rate.

More accurate estimates of the dependence of free-space \QKD\
performance on  source characteristics and on the communication
distance $d$ can be obtained by applying existing analyses of atmospheric
effects on optical signals~\cite{teich,teich2,gilbert}.

Finally, we mention a subtle issue in quantum cryptography that has
not, to our knowledge, been analyzed: the role of Alice's and Bob's
prior distribution on the error rate ($\ebar$) that Eve effects by her
eavesdropping.  In their attempts to determine $\ebar$, Alice and Bob
can only use the revealed outcome of a subset of the total
transmission record to update an {\em a priori} distribution over
$\ebar$ to an {\em a posteriori} distribution over $\ebar$ via Bayes'
rule.  While most practical analyses choose the uniform distribution
over $\ebar$ as the {\em a priori} distribution, Eve can obviously use
any distribution she likes to choose the value of $\ebar$.  Thus, it
seems likely that a more sophisticated game-theoretic analysis would be
required to plug this ``{\em a priori} distribution loophole.''

\section*{Acknowledgments}
Z.~W.~thanks M.~Du\v{s}ek and N.~L\"{u}tkenhaus for useful conversations.  This work has been supported by the National Science Foundation.

\end{document}